\documentclass[singlecolumn,showpacs,preprintnumbers,amsmath,amssymb,floatfix]{revtex4}
\usepackage{graphicx}% Include figure files
\usepackage{dcolumn}% Align table columns on decimal point
\usepackage{bm}% bold math

\begin{document}
%\preprint{APS/123-QED}

\title{Comparison of different proximity potentials for asymmetric colliding nuclei}
% Force line breaks with \\

\author{Ishwar Dutt }
%\altaffiliation [Also at ]{Physics Department, XYZ University.}%Lines break automatically or can be forced with \\
\author{Rajeev K. Puri}%
\email{rkpuri@pu.ac.in; drrkpuri@gmail.com}
\affiliation{Department of Physics, Panjab University, Chandigarh
160 014, India}

%\affiliation{

\date{\today}% It is always \today, today,

\begin{abstract}

Using the different versions of phenomenological proximity
potential as well as other parametrizations within the proximity
concept, we perform a detailed comparative study of fusion
barriers for asymmetric colliding nuclei with asymmetry parameter
as high as 0.23. In all, 12 different proximity potentials are
robust against the experimental data of 60 reactions. Our detailed
study reveals that the surface energy coefficient as well as
radius of the colliding nuclei depend significantly on the
asymmetry parameter. All models are able to explain the fusion
barrier heights within $\pm10\%$ on the average. The potentials
due to Bass 80, AW 95, and Denisov DP explain nicely the fusion
cross sections at above- as well as below-barrier energies.
\end{abstract}


% PACS, the Physics and Astronomy % Classification Scheme.
\maketitle

%break was forced \lowercase{via} \textbackslash\textbackslash}

\section{\label{intro}Introduction}
The fusion  of colliding nuclei with neutron -rich/ -deficient
content and at the extreme of isospin plane has attracted  a large
number of studies in recent
years~\cite{canto06,Aguilera95,Silva97,Aguilera90,Cavallera90,Vega90,Sonz98,Vinod96,Trotta01,Stefanini06,stefanini08}.
This renewed interest is due to the availability of
radioactive-ion beams that can produce nuclei at the extreme of
isospin~\cite{canto06,Stefanini06,stefanini08}. Further, this
field has also been enriched with several new phenomena  that put
a stringent test on theoretical models derived to study the fusion
phenomenon in heavy-ion reactions.
\par
As is evident from the literature, no experiment can extract
information about the fusion barriers directly. All experiments
measure the fusion differential cross
sections~\cite{canto06,Aguilera95,Silva97} and then with the help
of theoretical model, one extracts the fusion barriers.
Theoretical models are very helpful  in understanding  the nuclear
interactions at a microscopic level. A vast number of theoretical
models and potentials have become  available  in recent years that
can explain one or the other features of fusion
dynamics~\cite{id1,rkp1,rkp2,blocki77,wr94,ms2000,gr09,bass73,bass77,cw76,aw95,ngo80,deni02,ngo75,deni07}.
\par
In the galaxy of different theoretical models, proximity
potential~\cite{blocki77} enjoys very popular status. This
phenomenological potential is a benchmark and backbone for all
microscopic/macroscopic fusion models. It is almost mandatory to
compare the potential and parametrize it within the proximity
concept for wider acceptability. In recent years, several
refinements and modifications have been proposed over original
proximity potential~\cite{wr94,ms2000}. Further with the passage
of time, different versions of the same model are also
available~\cite{id1}. Many of these modifications are based on the
isospin effects either through the surface energy coefficients or
via nuclear radius. It would be of interest to test these
potentials in the isospin plane and to see how these different
potentials will perform when asymmetry in the neutron/proton
content is very large.
\par
Recently, we carried out a detailed comparative systematic study
 of different fusion models for symmetric colliding nuclei~\cite{id1}.
 Here we plan to extend this study for those colliding nuclei that have
larger neutron/proton content. In this study, we shall compare as
many as 12 proximity potentials with different versions. This will
include four versions of proximity potential, three versions of
potential due to Bass and Winther each and the latest potential
due to Ng\^{o} and a modified version of the Denisov potential.
Section~\ref{model}, deals with formalism in detail,
Sec.~\ref{result} contains the results, and a summary is presented
in Sec.~\ref{summary}.
%%%%%%%%%%%%%%%%%%%%%%%%%%%%%%%%%% Models %%%%%%%%%%%%%%%%%%%%%%%%%%%
\section{\label{model} Formalism}
In this section, we present the details of  various proximity
potentials
 used for the calculation of  fusion barriers. When two surfaces
 approach each other within a distance of 2 - 3 fm, additional
 force due to the proximity of the surface is labeled as
 proximity potential. Various versions of these potentials take
 care of different aspects including the isospin dependence. In the following, we discuss each of them in detail.
\subsection{$\rm Proximity~1977~(Prox~77)$}
   The basis of proximity
potential is the theorem that states that \emph{``the force
between two gently curved surfaces in close proximity is
proportional to the interaction potential per unit area between
the two flat surfaces''}. According to the original version of
proximity potential 1977~\cite{blocki77}, the interaction
potential $ V_{N}(r)$ between two
 surfaces can be written as:
 \begin{equation}
V_{N}(r)= 4\pi \gamma b \overline{R} \Phi \left( \frac{{ r}-C_{1}
-C_{2}}{b} \right) ~~\rm MeV. \label{eq:1}
\end{equation}
In this, the mean curvature radius, $ \overline{R}$ has the form
\begin{equation}
\overline{R} = \frac{C_{1}C_{2}}{C_{1}+ C_{2}}, \label{eq:2}
\end{equation}
quite similar to the one used for reduced mass. Here
\begin{equation}
C_{i}= R_{i}\left[1-\left(\frac{b}{R_{i}} \right)^{2}+\cdots
\cdots \right], \label{eq:3}
\end{equation}
${\rm R_{i}}$, the effective sharp radius, reads as
\begin{equation}
R_{i}= 1.28A^{1/3}_{i}- 0.76+0.8A^{-1/3}_{i} {~\rm
fm}~~~~~~~~~~~(i=1,2). \label{eq:4}
\end{equation}
In Eq.~(\ref{eq:1}), $\Phi(\xi = \frac{{ r}-C_{1} -C_{2}}{b})$ is
a universal function that depends on the separation between the
surfaces of two colliding nuclei only. As we see, both these
factors do not depend on the isospin content. However, $\gamma$,
the surface energy coefficient, depends on the neutron/proton
excess as
\begin{equation}
\gamma = \gamma_{0}\left[1-k_{s}\left(\frac{N-Z}{N+Z}\right)^{2}
\right],
 \label{eq:5}
\end{equation}
where N and Z are the total number of neutrons and protons. In the
present version, $\gamma_{0}$ and $k_{s}$ were taken to be
$0.9517~\rm MeV/fm^{2}$ and $1.7826$, respectively. Note that for
the symmetric colliding pair i.e. (N = Z),
$\gamma=\gamma_{0}=0.9517~\rm MeV/fm^{2}$. If the
$\left(\frac{N-Z}{N+Z}\right)$ ratio is $0.5$, $\gamma$ reduces to
$0.5276~\rm MeV/fm^{2}$. Defining asymmetry parameter
$A_{s}=\left[\frac{N_{1}+N_{2}-(Z_{1}+Z_{2})}{N_{1}+N_{2}+(Z_{1}+Z_{2})}\right]$,
one notices drastic reduction in the magnitude of the potential
with asymmetry of the colliding pair. Interestingly, most of the
modified proximity type potentials use different values of the
parameter $\gamma$~\cite{wr94,ms2000}.

 The universal function $\Phi \left(\xi
\right)$ was parameterized with the following form:
\begin{equation}
\Phi \left(\xi \right)= \left\{
\begin{array}{l r}
-\frac{1}{2} \left(\xi- 2.54 \right)^{2}-0.0852\left(\xi- 2.54 \right)^{3},            \\
         ~~~~~~~~~~~~~~~~~~~~~~~~~~~~~~~~ \mbox{ for $\xi \leq 1.2511 $ },    \\
-3.437\exp \left(-\xi/0.75 \right),          \\
 ~~~~~~~~~~~~~~~~~~~~~~~~~~~~~~~~\mbox{ for $\xi
\geq 1.2511 $ }.
\end{array}
\right. \label{eq:6}
\end{equation}
The surface width $b$ has been evaluated close to unity. Using the
above form, one can calculate the nuclear part of the interaction
potential ${ V_{N}(r)}$. This model is referred to as Prox 77 and
the corresponding potential as $ V_{N}^{Prox~77}(r)$.
\subsection{$\rm Proximity~1988~(Prox~88)$}
Later on, using the more refined mass  formula due to M\"oller and
Nix~\cite{mn81}, the value of coefficients $\gamma_{0}$ and $
k_{s}$ were modified yielding the values = 1.2496 $\rm MeV/fm^{2}$
and 2.3, respectively. Reisdorf \cite{wr94} labeled this modified
version
 as ``Proximity 1988". Note that this set of
 coefficients give stronger attraction  compared to the above set. Even a more recent compilation
 by M\"oller and  Nix \cite{mn95} yields similar results. We
labeled this potential as Prox 88.
\subsection{$\rm Proximity~2000~(Prox~00)$}
%\label{subsec:3.3.3} \hspace*{.5cm}
Recently, Myers and \'Swi\c{a}tecki~\cite{ms2000} modified
Eq.~(\ref{eq:1}) by using up-to-date knowledge of nuclear radii
and surface tension coefficients using their droplet model
concept. The prime aim behind this attempt was to remove
descripency of the order of $4\%$ reported between the results of
Prox 77 and experimental data~\cite{ms2000}.
 Using the droplet model
\cite{ms80}, matter radius $C_{i}$ was calculated as
\begin{equation}
C_{i}= c_{i}+ \frac{N_{i}}{A_{i}}t_{i}    ~~~~(i=1,2),
\label{eq:7}
\end{equation}
where $c_{i}$ denotes the half-density radii  of the charge
distribution and $t_{i}$ is the neutron skin of the nucleus. To
calculate $c_{i}$, these authors~\cite{ms2000} used  two-parameter
Fermi function values given in Ref. \cite{dv87} and the remaining
cases were handled with the help of parametrization of charge
distribution described below. The nuclear charge radius (denoted
as $R_{00}$ in Ref.~\cite{bn94}), is given by the relation:
\begin{equation}
R_{00i}= \sqrt{\frac{5}{3}}\left<r^{2}\right>^{1/2}
~~~~~~~~~~~~~~~~~~~~~~~~~~~~~~~~~~~~~~~~~~\nonumber
\end{equation}
\begin{eqnarray}
= 1.240A_{i}^{1/3} \left\{1+\frac{1.646}{A_{i}}
-0.191\left(\frac{A_{i}-2Z_{i}}{A_{i}}\right)\right\} {~\rm
fm} \nonumber\\
(i=1,2), \label{eq:8}
\end{eqnarray}
where $<r^{2}>$ represents the mean-square nuclear charge radius.
According to Ref.~\cite{bn94}, Eq.~(\ref{eq:8}) was valid for the
even-even nuclei with $8\leq Z < 38$ only. For nuclei with $Z\geq
38$, the above equation was modified by Pomorski \emph{et
al}.~\cite{bn94} as;
\begin{eqnarray}
R_{00i}=  1.256A_{i}^{1/3}
\left\{1-0.202\left(\frac{A_{i}-2Z_{i}}{A_{i}}\right)\right\}
{~\rm fm} \nonumber \\
~~~~~~(i=1,2). \label{eq:9}
\end{eqnarray}
These expressions give good estimate of  the measured mean square
nuclear charge radius $<r^{2}>$. In the present model, authors
used only  Eq.~(\ref{eq:8}). The half-density radius, $c_{i}$, was
obtained from the relation:
\begin{equation}
c_{i}= R_{00i}\left(1-
\frac{7}{2}\frac{b^{2}}{R_{00i}^{2}}-\frac{49}{8}\frac{b^{4}}{R_{00i}^{4}}+\cdots
\cdots   \right) ~~~~~~~(i=1,2). \label{eq:10}
\end{equation}
 Using the droplet
model~\cite{ms80}, neutron skin $t_{i}$ reads as;
\begin{equation}
t_{i}= \frac{3}{2}r_{0}\left[\frac{JI_{i}-
\frac{1}{12}c_{1}Z_{i}A^{-1/3}_{i}}{Q+ \frac{9}{4}JA^{-1/3}_{i}}
\right]     (i=1,2). \label{eq:11}
\end{equation}
Here $r_{0}$ is $1.14$ fm, the value of nuclear symmetric energy
coefficient $J=32.65$ MeV and $c_{1}= 3 e^{2}/5 r_{0}=0.757895$
MeV. The neutron skin stiffness coefficient Q was taken to be 35.4
MeV. The nuclear surface energy coefficient $\gamma$ in terms of
neutron skin was given as;
\begin{equation}
\gamma = \frac{1}{4\pi r^{2}_{0}}\left[18.63  {\rm  (MeV)}
-Q\frac{\left(t^{2}_{1} + t^{2}_{2}\right)}{2r^{2}_{0}} \right],
\label{eq:12}
\end{equation}
where $t_{1}$ and $t_{2}$ were calculated using Eq.~(\ref{eq:11}).
 The universal function $\Phi
(\xi)$ was reported as;
\begin{equation}
\Phi \left(\xi \right)=\left \{
\begin{array}{ll}
-0.1353+ \sum\limits_{n=0}^{5}\left[c_{n}/\left(n+1\right)\right] \left(2.5 - \xi \right)^{n+1},   \\
~~~~~~~~~~~~~~~~~~~~~~~~~~~~~~ \mbox{ for \quad $0 < \xi \leq  2.5 $},\\
-0.09551\exp \left[\left( 2.75 - \xi\right)/0.7176 \right], \\
~~~~~~~~~~~~~~~~~~~~~~~~~~~~~~\mbox{ for $\quad\xi \geq  2.5$}.
\end{array}
\right. \label{eq:13}
\end{equation}
The values of different constants $c_{n}$ were: $c_{0}=-0.1886$,
$c_{1}=-0.2628$, $c_{2}=-0.15216$, $c_{3}=-0.04562$,
$c_{4}=0.069136$ and $c_{5}=-0.011454$. For $\xi
> 2.74$, the above exponential expression is the exact representation of
the Thomas-Fermi extension of the proximity potential. This
potential is labeled Prox 00.
 \subsection{$\label{DP}\rm Modified ~Proximity~2000~(Prox~00DP)$}
Recently, Royer and Rousseau~\cite{gr09} modified Eq.~(\ref{eq:8})
with slightly different constants as;
\begin{eqnarray}
R_{00i}= 1.2332A^{1/3}_{i} \left[ 1+\frac{2.348443}{A_{i}}
\right.~~~~~~~~~~~~~~~~~~\nonumber \\
\left. -0.151541\left(\frac{A_{i}-2Z_{i}}{A_{i}}\right) \right]
 {~\rm fm}
~~(i=1,2). \label{eq:14}
\end{eqnarray}

It is obtained by analyzing as many as 2027 masses with N, Z
$\geq$ 8 and a mass uncertainty $\leq$ 150 keV. Further, the
accuracy of the above formula is mainly improved by adding the
Coulomb diffuseness correction or the charge exchange correction
to the mass formulas~\cite{gr09}. We implement this radius in the
proximity 2000 version  instead  of the form given in the
proximity 2000. This new version of the proximity potential is
labeled Prox 00DP~\cite{id1}.
 \subsection{$\rm Bass~1973~(Bass~73)$}
This model is based on the assumption of liquid drop model
~\cite{bass73}. Here change in the surface energy of  two
fragments due to their mutual separation is represented by
exponential factor. By multiply with geometrical arguments, one
can obtained the nuclear part of the interaction potential as
\begin{equation}
 V_{N}(r)^{Bass~73}= -
\frac{d}{R_{12}}a_{s}A_{1}^{1/3}A_{2}^{1/3}exp(-\frac{r-R_{12}}{d}){~\rm
MeV}, \label{eq:15}
\end{equation}
with  ${R_{12}}=r_{0}(A_{1}^{1/3}+A_{2}^{1/3}), ~d=1.35 ~{\rm fm}
$ and $a_{s}=17.0 ~{\rm MeV}$. The cut-off distance  $R_{12}$ is
chosen to yield saturation density in the overlap region and $
r_{o}=1.07 ~{\rm fm} $ corresponding half of the maximum density
for individual nucleus. We labeled this potential Bass 73.
 \subsection{$\rm Bass~1977~(Bass~77)$}
 In this model, nucleus-nucleus potential is derived from the
 information based on the experimental fusion cross sections  by
 using the liquid drop model and general geometrical arguments.
 The nuclear part of the potential (for spherical nuclei with frozen
 densities) can be written as~\cite{bass77}
 \begin{eqnarray}
V_{N}\left( r \right)^{Bass~77}= -4\pi \gamma \frac{R_1 R_2}{R_1 +
R_2} f\left(r-R_{1}-R_{2} \right)\nonumber\\  \qquad\qquad=
-\frac{R_1 R_2}{R_1 + R_2} \Phi\left(r-R_{1}-R_{2} \right)~{\rm
MeV}, \label{eq:16}
\end{eqnarray}
with
\begin{equation}
\frac{df}{ds}=-1, \qquad \qquad {\rm for} \qquad \qquad s = 0.
\label{eq:17}
\end{equation}
Note that $f\left(s=r-R_{1}-R_{2} \right)$ and
$\Phi\left(s=r-R_{1}-R_{2} \right)$ are the universal functions.
Here radius $R_{i}$ is written as
\begin{equation}
R_{i}= 1.16A^{1/3}_{i}-1.39A^{-1/3}_{i}  ~{\rm fm}~~~~(i=1,2).
\label{eq:18}
\end{equation}
The form of the universal function $\Phi\left(s \right)$ reads as
\begin{equation}
\Phi \left(s \right)= \left[A \exp \left( \frac{s}{d_1} \right) +
B\exp \left( \frac{s}{d_2}\right)\right]^{-1}, \label{eq:19}
\end{equation}
with $A= 0.0300$ MeV$^{-1}$fm, $B=0.0061$ MeV$^{-1}$fm, $d_1 =
3.30$ fm and $d_2 = 0.65$ fm. Note that where $b=1$, $\xi$ and s
turn out to be the same quantities. This model was very successful
in explaining the barrier heights, positions, and cross sections
over a wide range of incident energies and masses of colliding
nuclei. We labeled this potential Bass 77.
 \subsection{$\rm Bass~1980~(Bass~80)$}
The above potential form was further improved by Bass~\cite{wr94}.
Here $\Phi\left(s=r-R_{1}-R_{2} \right)$ is now  given as:
\begin{equation}
\Phi \left(s \right)= \left[0.033 \exp \left( \frac{s}{3.5}
\right) + 0.007\exp \left( \frac{s}{0.65}\right)\right]^{-1},
\label{eq:20}
\end{equation}
with central radius, $R_{i}$ as
\begin{equation}
R_{i}= R_{s}\left(1-\frac{0.98}{R_{s}^{2}} \right)
~~~~(i=1,2),\label{eq:21}
\end{equation}
where $R_{s}$ is same as given by Eq. (\ref{eq:4}). We labeled
this potential as Bass 80.
\subsection{$\rm Christensen ~ and ~Winther~1976~(CW~76)$}
Christensen and Winther ~\cite{cw76} derived the nucleus-nucleus
interaction potential by analyzing  the  heavy-ion
elastic-scattering data, based on the semiclassical arguments and
the recognition that optical-model analysis of elastic scattering
determines the real part of the interaction potential only in the
vicinity of a characteristic distance. The nuclear part of the
empirical potential due to Christensen and Winther  is written as
\begin{equation}
V_{N}^{CW~76}\left( r \right)= -50 \frac{R_1 R_2}{R_1 + R_2} \Phi
\left(r-R_{1}-R_{2} \right)~{\rm MeV}. \label{eq:22}
\end{equation}
This form of the geometrical factor is similar to that of $\rm
Bass~77$ with different radius parameters
\begin{equation}
R_{i}= 1.233A^{1/3}_{i} - 0.978A^{-1/3}_{i}~{\rm fm}~~~~(i=1,2).
\label{eq:23}
\end{equation}
The universal function $\Phi(s=r-R_{1}-R_{2}$ ) has the following
form
\begin{equation}
\Phi \left(s\right)= \exp \left( -\frac{r-R_{1}-R_{2}}{0.63}
\right). \label{eq:24}
\end{equation}
This model was tested for more than 60 reactions and we labeled it
CW 76.
\subsection{$\rm Broglia ~and ~Winther~1991~(BW~91)$}
A refined version of the above potential was derived by Broglia
and Winther~\cite{wr94}, by taking Woods-Saxon parametrization
with subsidiary condition of being compatible with the  value of
the maximum nuclear force predicted by the proximity potential
Prox 77. This refined potential resulted in
\begin{equation}
V_N^{BW~91}(r)=-\frac{V_{0}}{1+\exp
\left(\frac{r-R_0}{0.63}\right)} ~{\rm MeV}; \label{eq:25}
\end{equation}
\begin{equation}
{\rm with}, ~V_0=16\pi\frac{R_1 R_2}{R_1 + R_2}{\gamma} {a},
\label{eq:26}
\end{equation}
here $a=0.63$ fm and
\begin{equation}
R_0=R_1 + R_2 + 0.29. \label{eq:27}
\end{equation}
Here radius $R_{i}$ has the form
\begin{equation}
R_{i}= 1.233A^{1/3}_{i} - 0.98A^{-1/3}_{i}~{\rm fm}~~~~(i=1,2).
 \label{eq:28}
\end{equation}
The form of the surface energy coefficient $\gamma$ is  quite
similar to the one used in Prox 77 with slight difference
 \begin{equation}
\gamma = \gamma_{o}\left[1-k_{s}
\left(\frac{N_{p}-Z_{p}}{A_{p}}\right)\left(\frac{N_{t}-Z_{t}}{A_{t}}\right)
\right],
 \label{eq:29}
\end{equation}
where ${~\rm \gamma_{0}}$ = 0.95 $~\rm
 ~MeV/fm^{2}$ and $ k_{s}=1.8$. Note that the second term used in this potential gives different
results when the projectile is symmetric ($N = Z$) and the target
is asymmetric ($N > Z$).  This form will also give different
results for larger mass asymmetry $\eta_{A}$. Note that the radius
used in this potential has same form like that of Bass with
different constants. We labeled this potential as BW 91.
\subsection{$\rm Aage~Winther~(AW~95)$}
Winther adjusted the parameters of the above potential through an
extensive comparison with experimental data for heavy-ion elastic
scattering. This refined adjustment to slight different values of
``$a$" and $R_{i}$ as~\cite{aw95}
\begin{equation}
{a}=\left[\frac{1}{1.17(1+0.53(A_{1}^{-1/3}+A_{2}^{-1/3}))}\right]
~\rm fm, \label{eq:30}
\end{equation}
and
\begin{equation}
R_{i}= 1.20A^{1/3}_{i} - 0.09 ~\rm fm~~~~(i=1,2).
 \label{eq:31}
\end{equation}
Here, $R_{0}=R_{1}+R_{2}$ only. We labeled this potential as AW
95.

\subsection{$\rm Ng$\^o$~1980~(\rm Ng$\^o~$80)$}
In earlier attempts, based on the microscopic picture of a nucleus
and on the idea of energy density formalism, the potential from
Ng\^o and collaborators enjoy special status~\cite{ngo75}. In this
model, calculations of the ion-ion potential are performed within
the framework of energy density formalism due to Bruckener
\emph{et al}., using a sudden approximation~\cite{bk68}. The need
of Hartree-Fock densities as input in this model limited its
scope. This not only made calculations tedious, but it also
hindered its application to heavier nuclei. The above-stated
parametrization was improved by H. Ng\^o and Ch.
Ng\^o~\cite{ngo80}, by using a Fermi-density distribution for
nuclear densities as
\begin{equation}
\rho_{n, p} (r)= \frac{\rho_{n, p}(0)}{1+\exp\left[(r-C_{n,
p})/0.55\right]}~, \label{eq:34}
\end{equation}
where $C$ represents the central radius of the distribution and is
defined in Prox 77~(see Eq. (\ref{eq:3}) with b = 1 fm). Here
$\rho_{n, p}(0)$ is given by
\begin{equation}
\rho_n(0)=\frac{3}{4\pi}\frac{N}{A}\frac{1}{r^3_{0_{n}}};~~~~~~~\rho_p(0)=\frac{3}{4\pi}\frac{Z}{A}\frac{1}{r^3_{0_{p}}}~.
\label{eq:35}
\end{equation}
 Ng\^o parameterized the nucleus-nucleus interaction
potential in the spirit of proximity concept. The interaction
potential can be divided into the geometrical factor and a
universal function. The nuclear part of the parameterized
potential is written as~\cite{ngo80};
\begin{equation}
V_{N}^{Ngo~80}\left( r \right)=  \overline{R} \Phi \left(r - C_{1}
- C_{2} \right)~\rm MeV, \label{eq:36}
\end{equation}
where $\overline{R}$ is defined by Eq.~(\ref{eq:2}). Now the
nuclear radius $R_{i}$ reads as:
\begin{equation}
R_{i}= \frac{NR_{n_{i}} +ZR_{p_{i}}}{A_{i}}~~~~~(i=1,2).
\label{eq:37}
\end{equation}
The equivalent sharp radius for protons and neutrons are given as;
\begin{equation}
R_{p_{i}}= r_{0_{pi}}A^{1/3}_{i};~~~~~~~~~~~~~R_{n_{i}}=
r_{0_{ni}}A^{1/3}_{i}, \label{eq:38}
\end{equation}
with
\begin{equation}
 r_{0_{pi}}= 1.128~{\rm fm};~ r_{0_{ni}}= 1.1375 +
1.875\times 10^{-4} A_{i}~\rm fm. \label{eq:39}
\end{equation}
The above different radius formulas for the neutrons and protons
take isotopic dependence into account. The universal function
$\Phi (s=r - C_{1} - C_{2})$ (in $\rm MeV/ fm$) is noted by
\begin{equation}
\Phi \left(s \right)= \left\{
\begin{array}{l r}
-33 + 5.4\left(s- s_{0} \right)^{2},                     & \mbox{ for $s < s_{0} $ },    \\
-33\exp \left[-\frac{1}{5}\left(s-s_{0}\right)^{2} \right], &
\mbox{ for $s \ge  s_{0}$ },
\end{array}
\right. \label{eq:40}
\end{equation}
and $s_{0} = -1.6$ fm. We labeled this potential as Ng\^o 80.
\subsection{\rm New Denisov Potential (Denisov DP)}
\label{subsec:4.3.1}
\par
Denisov~\cite{deni02} performed numerical calculations and
parametrized the potential based on 7140 pair within
semi-microscopic approximation. In total, 119 spherical or near
spherical nuclei along the $\beta$-stability line from $^{16}$O to
$^{212}$Po were taken. The potential is evaluated for any
nucleus-nucleus combinations at 15 distances between ions around
the touching point. By using this database, a simple analytical
expression for the nuclear part of the interaction potential
V$_{N}$(r) between two spherical nuclei is presented as;
\begin{eqnarray}
 V_{N}\left(r \right)  =  -1.989843\frac{R_{1}R_{2}}{R_{1}+ R_{2}} \Phi \left(r - R_{1} - R_{2}-2.65 \right)      \nonumber    \\
  \times  \left[1+0.003525139\left(\frac{A_{1}}{A_{2}}+ \frac{A_{2}}{A_{1}} \right)^{3/2} \right. \nonumber  \\
\left. -0.4113263\left(I_{1}+I_{2} \right) \right],
 \label{eq:41}
\end{eqnarray}
with
\begin{equation}
 I_{i} = \frac{N_{i}-Z_{i}}{A_{i}}~~~~(i=1,2).
 \label{eq:42}
\end{equation}
The effective nuclear radius $R_{i}$ is given as;
\begin{eqnarray}
 R_{i}= R_{i p}\left(1- \frac{3.413817}{R^{2}_{i p}} \right) + ~~~~~~~~~~~\nonumber\\
 1.284589\left(I_{i}- \frac{0.4A_{i}}{A_{i}+200}
\right) (i=1,2),
 \label{eq:43}
\end{eqnarray}
where, proton radius $R_{i p}$ is given by Eq.~(\ref{eq:8}) and
$\Phi \left(s= r - R_{1} - R_{2}-2.65 \right)$ is given by the
following complex form:
\begin{equation}
 \Phi (s) = \left\{
\begin{array}{l}
1- s/0.7881663  +1.229218s^{2} -0.2234277s^{3}\\
 -0.1038769s^{4}\\
 -\frac{R_{1}R_{2}}{R_{1}+ R_{2}}
\left(0.1844935s^{2}+0.07570101s^{3} \right)\\
+ \left(I_{1}+I_{2} \right) \left(0.04470645s^{2}+0.03346870s^{3} \right),     \\
    \qquad\qquad\qquad\qquad  {\rm for} \quad \quad -5.65\leq s\leq    0,\\
\left[1-s^{2} [0.05410106 \frac{R_{1}R_{2}}{R_{1}+ R_{2}}
\exp (-\frac{s}{1.760580}) \right. \\
\left. -0.5395420(I_{1}+I_{2}) \exp(-\frac{s}{2.424408})] \right] \\
\times \exp(-\frac{s}{0.7881663}), \\
\qquad \qquad \qquad\qquad \qquad \qquad {\rm for}\qquad  s\geq 0.
\end{array}
\right.
 \label{eq:44}
\end{equation}
Here $A_{i}$, $N_{i}$, $Z_{i}$, $R_{i}$, and $R_{i p}$ are,
respectively, the mass number, the number of neutrons, the number
of protons, the effective nuclear radius, and the proton radius of
the target and projectile. The above form of the universal
function not only depends on the separation distance s, but also
has complex dependence on the mass as well as on the relative
neutron excess content. The above parametrization is derived for
different combinations of nuclei between $^{16}$O and $^{212}$Po.
\par
As stated in the subsection~\ref{DP}, a new radius formula has
become available recently~\cite{gr09}. We here extend the above
potential due to Denisov to include this radius in its
parametrization. This modified new version of the potential is
labeled as Denisov DP~\cite{id1}. Note that this new
implementation was
 reported to yield very close agreement~(within 1\%) with experimental data for symmetric colliding pairs~\cite{id1}.

\par
If one looks on the different versions of potentials (Bass 73,
Bass77, Bass 80, and CW 76), one notices that although the form of
the radius is different, it is still isospin independent. Further,
the corresponding universal functions are also isospin
independent. The newer versions of Winther (BW 91 and AW 95) have
incorporated a $\gamma$ similar to the one used in the Prox 77
potential with a slightly different form. Here isospin content is
calculated separately for the target/ projectile. The latest
version of Ng\^{o}~(Ng\^{o} 80) has some isospin dependence in the
radius parameter. In most of the above mentioned potentials,
modifications are made either through the surface energy
coefficients or via nuclear radii. Both of these technical
parameters can have sizable effects on the outcome of a
reaction~\cite{id2}.
\par
 Using the above sets of models, the nuclear part of the
 interaction potential is calculated.

  By adding the Coulomb potential to a nuclear part, one can compute the total
 potential $V_{T}(r)$ for spherical colliding pairs as
 \begin{eqnarray}
 V_{T}(r)= V_{N}(r) + V_{C}(r),\\
         =V_{N}(r) + \frac{Z_{1}Z_{2}e^{2}}{r}.
\label{eq:30}
\end{eqnarray}
 Since the fusion
happens at a distance larger than the touching configuration of
colliding pair, the above form of the Coulomb potential is
justified. One can extract the barrier height $V^{theor}_{B}$ and
barrier position $R^{theor}_{B}$ using the following conditions
\begin{equation}
\frac{dV_T(r)}{dr}|_{r = R^{theor}_{B}} = 0;~~ {\rm{and}} ~~
\frac{d^{2}V_T(r)}{dr^{2}}|_{r = R^{theor}_{B}} \leq 0.
\label{eq:31}
\end{equation}
The knowledge of the shape of the potential as well as barrier
position and height, allows one to calculate the fusion
cross section at a microscopic level. To study the fusion cross
sections, we shall use the model given by Wong ~\cite{wg72}. In
this formalism, the cross section for complete fusion is given by
\begin{equation}
\sigma _{fus}= \frac{\pi}{k^{2}}\sum _{l=0}^{ l_{max}} \left(2l+1
\right)T_{l}\left(E_{cm} \right), \label{eq:32}
\end{equation}
where $k= \sqrt{\frac{2 \mu E}{\hbar^{2}}}$ and here $\mu$ is the
reduced mass. The center-of-mass energy is denoted by $E_{cm}$. In
the above formula, $\l_{max}$ corresponds to the largest partial
wave for which a pocket still exists in the interaction potential
and T$_{\l}\left(E_{cm} \right)$ is the energy-dependent barrier
penetration factor and is given by,
\begin{equation}
T_{\l}\left(E_{cm} \right)= \left\{1+ \exp \left[ \frac{2
\pi}{\hbar \omega_{\l}} \left(V^{theor}_{B_{\l}} - E_{cm}
\right)\right] \right\}^{-1},
 \label{eq:33}
\end{equation}
where $\hbar\omega_{l}$ is the curvature of the inverted parabola.
If we assume that the barrier position and width are independent
of $\l$, the fusion cross section reduces to
\begin{eqnarray}
\sigma _{fus}(mb)= \frac{10 R^{theor^{2}}_{B}\hbar \omega_{0}
}{2E_{cm}}\times\nonumber~~~~~~~~~~~~~~~~~~~~~~~~~~~~~~~~~ \\
 \ln \left\{1+ \exp\left[\frac{2\pi}{\hbar \omega _{0}}
\left(E_{cm}-V^{theor}_{B} \right)\right] \right\}. \label{eq:34}
\end{eqnarray}
For E$_{cm}$$>>$V$^{theor}_{B}$, the above formula reduces to
well-known sharp cut-off formula
\begin{equation}
\sigma _{fus}(mb)= 10 \pi R^{theor^{2}}_{B} \left(1 -
\frac{V^{theor}_{B}}{E_{cm}} \right), \label{eq:35}
\end{equation}
whereas for E$_{cm}$$<<$V$^{theor}_{B}$, the above formula reduces
to
\begin{equation}
\sigma _{fus}(mb)= \frac{10 R^{theor^{2}}_{B}\hbar \omega_{0}
}{2E_{cm}}\exp\left[\frac{2\pi}{\hbar \omega _{0}}
\left(E_{cm}-V^{theor}_{B} \right)\right]. \label{eq:36}
\end{equation}
We used Eq.~(\ref{eq:34}) to calculate the fusion cross sections.

\par
From the above brief discussion, it is clear that the main stress
is made on the surface energy coefficients $\gamma$ and nuclear
radii to incorporate the isospin dependence in the nuclear
potential. Definitely, the response of the isospin dependent
potentials will be different for asymmetric nuclei compared to
symmetric nuclei. At intermediate energies, a strong effect was
reported for the asymmetric reactions as well as for the mass
dependence of the reaction~\cite{rkp}.
%%%%%%%%%%%%%%%%%%%%%%%%%%%%%%%%%%%%%%%%%%%%%%%%%%%%%%X  Table-1 X%%%%%%%%%%%%%%%%%%%%%%%%%%%%%%%%%%%%%%%%%%%%%%%%%%%%%%%%%5

\section{\label{result}Results and Discussions}
The present study is conducted using a variety of the
above-mentioned potentials. In total, 60 asymmetric reactions with
compound mass between 29 and 294 (that have been experimentally
explored) are taken for the present study. All nuclei considered
here are assumed to be spherical in nature; however, deformation
as well as orientation of the nuclei also affect the fusion
barriers~\cite{deni07}. For uniform comparison of different
models, we consider all colliding nuclei to be spherical. The
lightest reaction taken is that of $^{12}$C + $^{17}$O, whereas
heaviest one is of $^{86}$Kr + $^{208}$Pb. The asymmetry $A_{s}$
of the colliding nuclei varies between 0.02 and 0.23. The other
form of the asymmetry used in the literature is the mass asymmetry
$\eta_{A}$~\cite{rkp1,rkp2}. In the present analysis, $\eta_{A}$
varies between 0.0 and 0.97. Note that the non zero value of
$A_{s}$ will involve complex interplay of the isospin degree of
freedom which has strong role at intermediate energies as well.
The variation of $\eta$ alters the physical outcome of a reaction
with $\eta \approx 0.0$ leading to high dense matter and maximum
collision volume whereas a larger value of $\eta \approx 1.0$ will
not be able to compress the matter to higher density~\cite{rkp}.
  %%%%%%%%%%%%%%%%%%%%%%%%%%%%%%%%%%Figure 5 %%%%%%%%%%%%%%%%%%%%%%%%%%

\par
As stated above, the isospin dependence of the different
potentials enters via surface energy coefficient $\gamma$. In Fig.
1, we display the variation of $\gamma$ (in $\rm MeV ~fm^{-2}$)
with asymmetry parameter $A_{s}$. Here we compare three versions
of the surface energy coefficient $\gamma$ used in Prox 77, Prox
88, and Prox 00 potentials along with the relation suggested in AW
95 potential. For the present analysis, the mass of the reacting
partner is kept fixed equal to $A_{1} = A_{2} = 40$. The $A_{s}$
was increased by increasing the neutrons and decreasing the
protons. For example, $^{40}_{20}Ca_{20}$ + $^{40}_{20}Ca_{20}$
has $A_{s}$ = 0.0. For $A_{s}$ = 0.2, we chose the reaction of
$^{40}_{16}S_{24}$ + $^{40}_{16}S_{24}$ whereas for $A_{s}$ = 0.4,
the reaction was $^{40}_{12}Mg_{28}$ + $^{40}_{12}Mg_{28}$. In all
cases, the mass of the reacting partner is kept fixed, whereas the
ratio $A_{s}$ is varied by converting the proton into neutrons. At
the end of this series, we have the reaction of
$^{40}_{10}Ne_{30}$ + $^{40}_{10}Ne_{30}$ having $A_{s}$ = 0.5.
 From the figure, we see that the surface energy coefficient
$\gamma$ used in the latest proximity potentials Prox 00/ Prox
00DP as well as in original version Prox 77 is less sensitive
toward the asymmetry and isospin dependence, whereas the one used
in the Prox 88 potential has a stronger dependence on the
asymmetry of the reacting nuclei. The coefficient $\gamma$ of AW
95 yields same results like Prox 77. Since nuclear potential
$V_{N}(r)$ depends directly on $\gamma$, one can conclude that the
potentials calculated within Prox 88 and Prox 77 will be far less
attractive for larger asymmetries compared to the one generated
using Prox 00. When colliding nuclei are symmetric (N = Z; $A_{s}$
= 0.0), such dependence does not play a role. In many
studies~\cite{rkp1}, one finds that neutron excess, leads to more
attraction. In these studies, the total mass of the colliding pair
is not fixed and as a result, this dependence is more of mass
dependence than of isospin dependence.

\par
In Fig. 2, we display the dependence of different nuclear radii on
the asymmetry  parameter $A_{s}$. As noted above, this parameter
also plays significant role in nuclear potential and finally in
the barrier calculations. We show the dependence of different
forms of nuclear radii used in various potentials on the asymmetry
parameter. We see that, the radius used in the Prox 77 (also in
Prox 88) as well as in Bass versions (i.e., Bass 73, Bass 77, and
Bass 80), and in all versions from Winther (CW 76, BW 91, and AW
95) are independent of the asymmetry content, whereas, the one
used in the Prox 00, Prox 00DP (and Denisov DP), and Ng\^o 80
versions depends on the asymmetry content of the colliding pairs.
\par
 From Figs. 1 and 2, we see that both these
parameters can lead to significant change in the nuclear potential
and ultimately in the fusion barriers  even if the universal
function $\Phi$(s) is kept the same.
\par
 In Fig. 3, we display the
nuclear part of the interaction potential $V_{N}(r)$ at a distance
of $C_{1}+C_{2}+1$ fm for the same sets of the reactions as
depicted in Figs. 1 and 2. In addition, a series of heavier
reacting partners with mass $A_{1}=A_{2}=80$ is also taken. We
display four versions of proximity potential, three versions from
Bass and Winther and one each of the latest versions of Ng\^o and
Denisov each. We see a systematic decrease in the attractive
strength of the potentials with asymmetry content $A_{s}$. The
decrease is stronger for the Prox 88 version compared to Prox 77,
Prox 00, and Prox 00DP. The Bass 73, Bass 77, and Bass 80 versions
of the potential are independent of the asymmetry content. One
also notices a very weak dependence in the Ng\^o 80 potential. Two
of the three versions of Winther potential have significant
dependence on the asymmetry of the reaction. The Winther 1976
potential, however, does not show such dependence due to the
absence of $\gamma$ term in the potential. The Denisov DP
potential also shows a linear decrease in the strength of the
potential with asymmetry content. These variations are stronger
for heavier colliding nuclei. This figure shows true isospin
dependence of the nuclear potential as the mass of the colliding
nuclei is kept fixed. All those potentials that do not depend on
the asymmetry parameter $A_{s}$ will not show any change in the
structure.

%%%%%%%%%%%%%%%%%%%%%%%%%%%%%%%%%%%%%%%%%%%%Table-1%%%%%%%%%%%%%%%%%%%%%%%%

\par
We now shift from the systematic study to the study involving real
nuclei. As stated above, here 60 reactions with $A_{s}$ between
0.02 and 0.23 and $\eta_{A}$ between  0.0 and 0.97 are taken. For
all these reactions, experimental fusion barriers are
known~\cite{Aguilera95,Silva97,Aguilera90,Cavallera90,Vega90,Sonz98,Vinod96,Trotta01,Stefanini06,stefanini08,Padron02,tighe93,vaz81,beck03,rath09,kolata98,Morsad90,trotta2000,gomes91,newton04,Liu05,Kolata04,Prasad96,Sinha01,Szanto90,Stefanini02,quirz01,Stefanini2000,Baby2000,capurro02,Vandana01,Stelson90,Mitsuoka07}.
In Fig. 4, we display the fusion barrier heights $V_{B}$ and
barrier positions $R_{B}$ versus experimental values for the above
mentioned reactions involving 12 different potentials. For the
clarity of the figure, only 60 asymmetric reactions studied
experimentally and covering the whole range of the mass and
asymmetry are displayed. We see no clear difference with fusion
barrier heights and positions. The fusion barrier heights can be
reproduced within $\pm10\%$ in all cases on the average. Due to
the large uncertainty in the fusion barrier positions, no definite
trend and conclusion can be drawn as is observed for the symmetric
colliding nuclei~\cite{id1}. To further understand the role of
isospin content, we display, in Fig. 5, the percentage difference
of the fusion barrier heights $\Delta V_{B} (\%)$ defined as
\begin{equation}
\Delta V_{B}~(\%) = \frac{V_{B}^{theor} -
V_{B}^{expt}}{V_{B}^{expt}}\times 100, \label{eq:17}
\end{equation}
verses asymmetry parameter $A_{s}$. In some cases, only the latest
versions of the potential are shown. Interestingly, we see that
Prox 77 and Ng\^o 80 fail to reproduce  the barrier heights
satisfactorily, whereas Prox 88, Bass 80, AW 95, Prox 00DP, and
Denisov DP do a far better job compared to other potentials. We do
not see any systematic deviation/improvement in the fusion barrier
heights with the asymmetry of the colliding nuclei. We see that
the potentials Prox 88, Bass 80, AW 95, and Denisov DP can
reproduce the empirical barrier heights within $\pm5\%$ (see the
shaded regions in Fig. 5), whereas others need $\pm10\%$ to
produce the same result.
\par
The comparison of the fusion barrier positions outcome is shown in
Fig. 6. We see that due to large uncertainty in the measurements
of fusion barrier positions, a large deviation is seen and all the
models are able to reproduce the results within $\pm10\%$. The
precise values of the fusion barrier heights $V_{B}$ (in MeV) and
positions $R_{B}$ (in fm) are shown in the Tables \ref{table} and
\ref{table1}, for 60 asymmetric colliding nuclei involving
significant variations of asymmetry $A_{s}$ as well as mass
asymmetry $\eta_{A}$. The experimental (or empirical) barriers
displayed in Tables \ref{table} and \ref{table1} and in Figs. 4 -
6 are obtained by fitting the cross sections in the approach, when
shapes of both colliding nuclei are spherical. A large number of
experimental data are available for different reactions; however,
we restrict ourselves to the latest one only.
\par
In Figs. 7 and 8, we display the fusion cross-sections
$\sigma_{fus}$ (in mb) as a function of center-of-mass energy
$E_{cm}$ for the reactions of $^{48}$Ca +
$^{96}$Zr~\cite{Stefanini06}, $^{28}$Si +
$^{92}$Zr~\cite{newton01}, $^{12}$C + $^{92}$Zr~\cite{newton01},
$^{16}$O + $^{208}$Pb~\cite{Morton01}~(in Fig. 7) and $^{16}$O +
$^{50}$Ti~\cite{Neto90}, $^{16}$O + $^{112}$Sn~\cite{Vandana01},
$^{16}$O + $^{116}$Sn~\cite{Vandana01}, and $^{16}$O +
$^{120}$Sn~\cite{Baby2000}~(in Fig. 8). Here the latest versions
of proximity parametrizations along with original proximity
potential and its modifications are shown for clarity.  As we see,
Bass 80, Denisov DP, and AW 95 do a better job for all the
systems, whereas Prox 77 and Ng\^o 80 fail to come closer to the
experimental data. The above results are in agreement with the one
obtained for symmetric colliding nuclei~\cite{id1}.
%%%%%%%%%%%%%%%%%%%%%%%%%%%%%%%%%%%%%%%%%%%%%%%X Summary X%%%%%%%%%%%%%%%%%%%%%%%%%%%%%%%
\section{\label{summary}Summary}
We performed a systematic study of the role of isospin dependence
on fusion barriers
  by employing as many as 12 different proximity-based potentials.
  Some of the potentials have isospin dependence via the surface energy
 coefficient as well as via nuclear radius. We
 noted that the nuclear part of the
 potential becomes more shallow with asymmetry of the reaction. On the
 other hand, a detailed comparison of different potentials does not
 show any preference for the isospin-dependent potential. Our comparison for 60 reactions reveals that all models can explain the fusion barrier heights within $\pm10\%$.
 The potentials from Prox 88, Bass 80, AW 95, and Denisov DP perform better than others. The fusion cross sections are nicely explained by Bass 80, AW 95, and Denisov DP
 potentials at below as well as above barrier energies.

This work was supported by a research grant from the Department of
Atomic Energy, Government of India.
%%%%%%%%%%%%%%%%%%%%%%%%%%%%%%%%%%%%%%%%%%%%%%%%%%%%%%%%%%%%%%%
%%%%%%%%%%%%%%%%%%%%%%%%%%%%%%%%%%%%%%%%%%%%%%%%%%%%%%%%%%%%%%%%%%%%%%%%%%%%%%%%%%%%%%%%%%

\newpage
%%%%%%%%%%%%%%%%%%%%%%%%%%%%%%%%%%%%%%%%%%%%%%%%%%%%%%%%%%%
\begin{figure}[!t]
\begin{center}
\includegraphics*[scale=0.4] {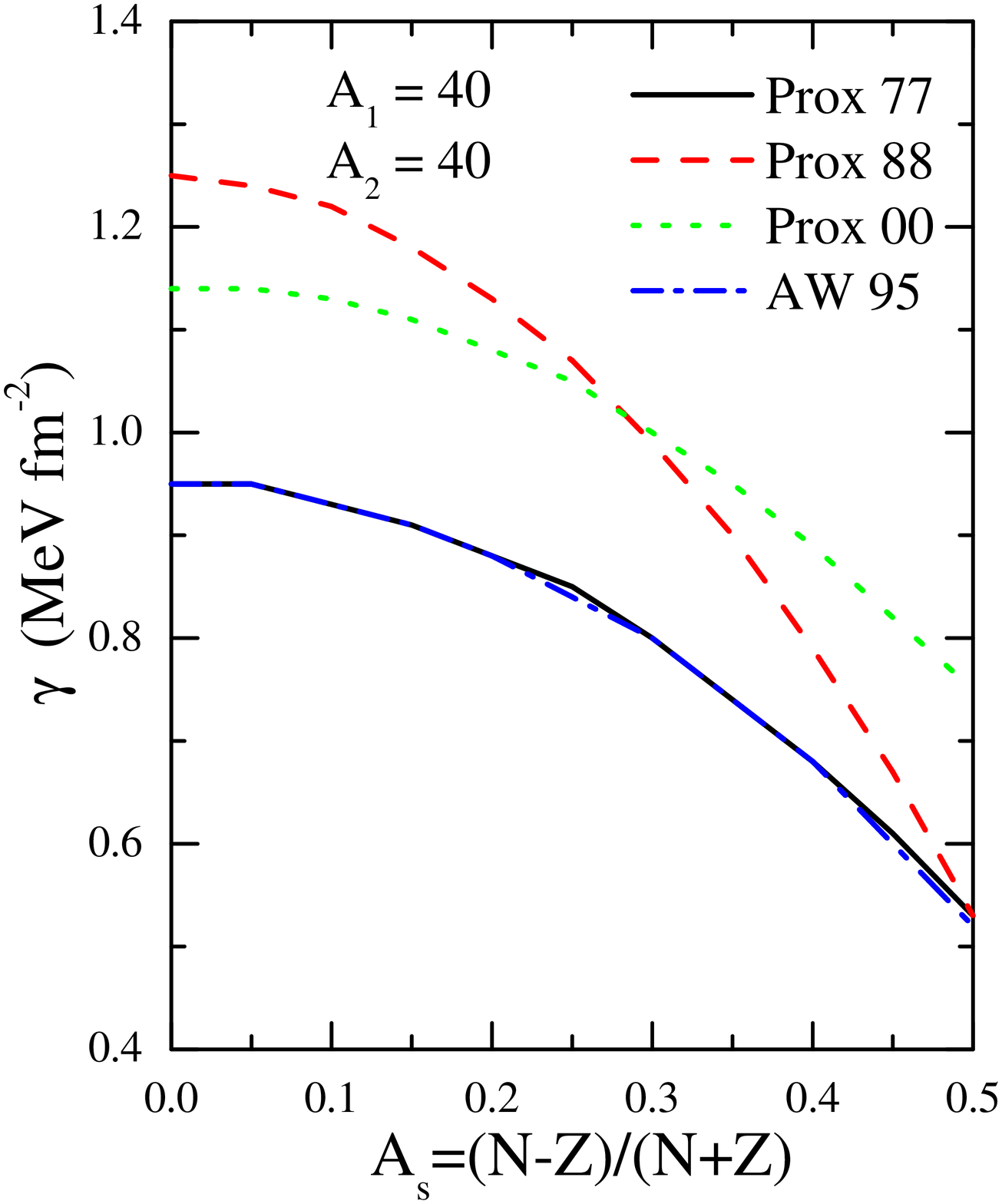}% Here is how to import EPS art
\vskip -0.50 cm \caption {(Color online) The variation of the
surface energy coefficient $\gamma$ $(\rm MeV ~fm^{-2})$ with
asymmetry parameter $A_{s}$. We display the results using $\gamma$
from Prox 77, Prox 88, Prox 00, and AW 95 for masses of reacting
partner $A_{1}$ = $A_{2}$ = 40 units. }
\end{center}
%\vskip -4.4 cm
\end{figure}
%%%%%%%%%%%%%%%%%%%%%%%%%%%%%%%%%Figure 2 %%%%%%%%%%%%%%%%%%%%%%%%%%
\begin{figure}[!t]
\begin{center}
\includegraphics*[scale=0.4] {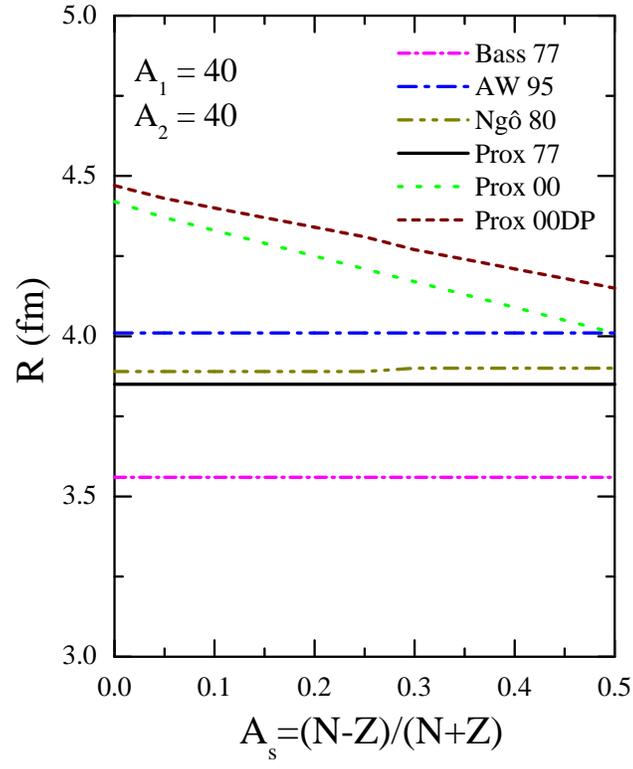}% Here is how to import EPS art
\vskip -0.50 cm \caption {(Color online) Same as Fig. 1, but for
various radii used in the literature.}
\end{center}
%\vskip -4.4 cm
\end{figure}
%%%%%%%%%%%%%%%%%%%%%%%%%%%%%%%%%%%%%%%%%%%%%%%%%%%%%%%%%%%%%%%%%%%%%%%%%%%%%%%%%%%%%%%%%%
     %%%%%%%%%%%%%%%%%%%%%%%%%%%%%%%%%%Figure 3 %%%%%%%%%%%%%%%%%%%%%%%%%%
\begin{figure}[!t]
\begin{center}
\includegraphics*[scale=0.4] {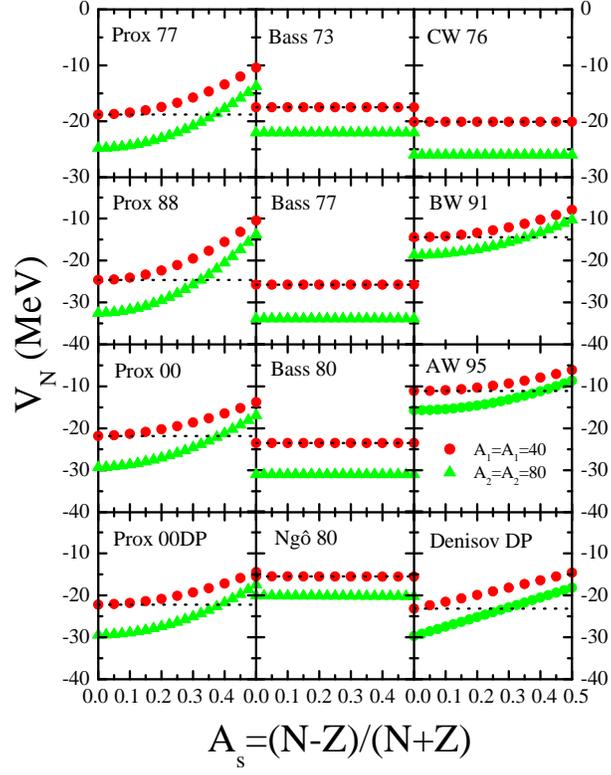}% Here is how to import EPS art
\vskip -0.5 cm \caption {(Color online) The strength of the
nuclear potential $\rm V_{N}~ (MeV)$ calculated at a distance
equal to $C_{1}+C_{2}+1$ fm as a function of asymmetry
 parameter $A_{s}$ for the reacting partners having masses $A_{1}$ = $A_{2}$ = 40 and $A_{1}$ = $A_{2}$ = 80 units.
 Here $C_{i}$ denotes the central radius~\cite{id1}. The dotted lines denote the
 value of the potential at $A_{s}$ = 0.0 (for $A_{1}$ = $A_{2}$ = 40 only) using proximity potentials. }
\end{center}
%\vskip -4.4 cm
\end{figure}
%%%%%%%%%%%%%%%%%%%%%%%%%%%%%%%%%%%%%%%%%%%%%%%%%%%%%%%%%%%%%%%%%%%%%%%%%%%%%%%%%%%%%%%%%%

 %%%%%%%%%%%%%%%%%%%%%%%%%%%%%%%%%%Figure 4 %%%%%%%%%%%%%%%%%%%%%%%%%%
\begin{figure}[!t]
\begin{center}
\includegraphics*[scale=0.4] {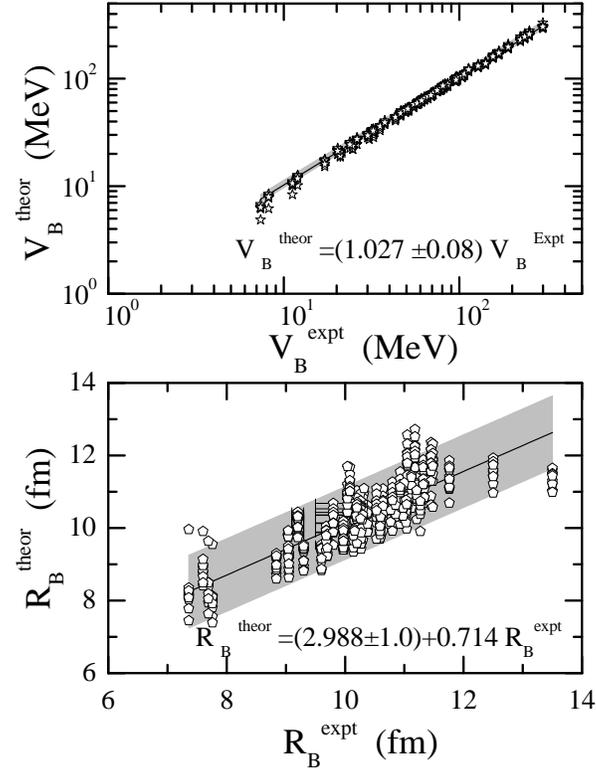}% Here is how to import EPS art
\vskip -0.50 cm \caption {The theoretical fusion barrier heights
$V_{B}$ (MeV) and positions $R_{B}$ (fm) are displayed as a
function of experimentally extracted values. The shaded area
represents the region within which all 12 proximity potentials are
able to reproduce experimental data.}
\end{center}
%\vskip -4.4 cm
\end{figure}
%%%%%%%%%%%%%%%%%%%%%%%%%%%%%%%%%%%Results and discussion%%%%%%%%%%%%%%%%%%%%%%%%%%%%%%%%%%
\begin{figure}[!t]
\begin{center}
\includegraphics*[scale=0.4] {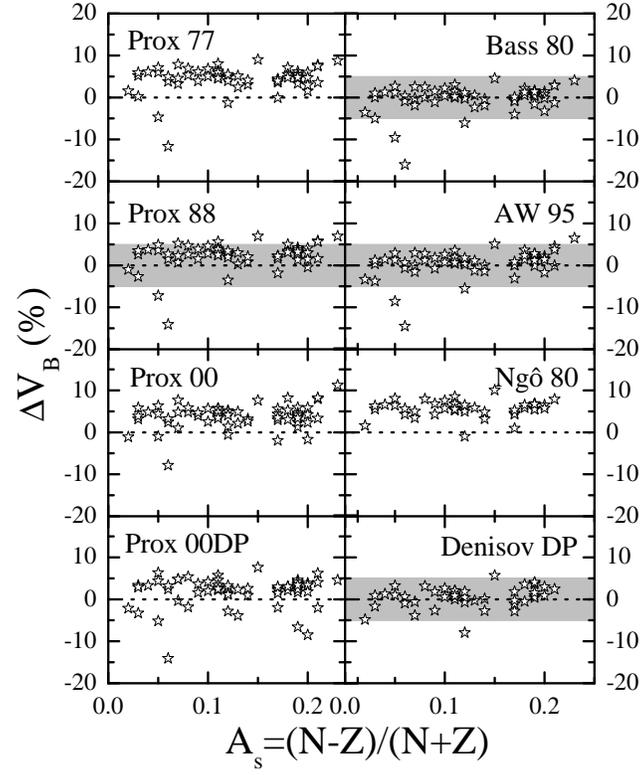}% Here is how to import EPS art
\vskip -0.50 cm \caption {The percentage difference $\Delta V_{B}
(\%)$ of theoretical fusion barrier heights over experimental
 one as a function of asymmetry parameter $A_{s}$.
 Here only 60 reactions covering the whole mass and asymmetry range are
taken. The shaded area is marked only for those potentials where
the deviation is within $\pm 5\%$. }
\end{center}
%\vskip -4.4 cm
\end{figure}
%%%%%%%%%%%%%%%%%%%%%%%%%%%%%%%%%%%%%%%%%%%%%%%%%%%%%%%%%%%%%%%%%%%%%%%%%%%%%%%%%%%%%%%%%%

%%%%%%%%%%%%%%%%%%%%%%%%%%%%%%%%%%Figure 6 %%%%%%%%%%%%%%%%%%%%%%%%%%
\begin{figure}[!t]
\begin{center}
\includegraphics*[scale=0.4] {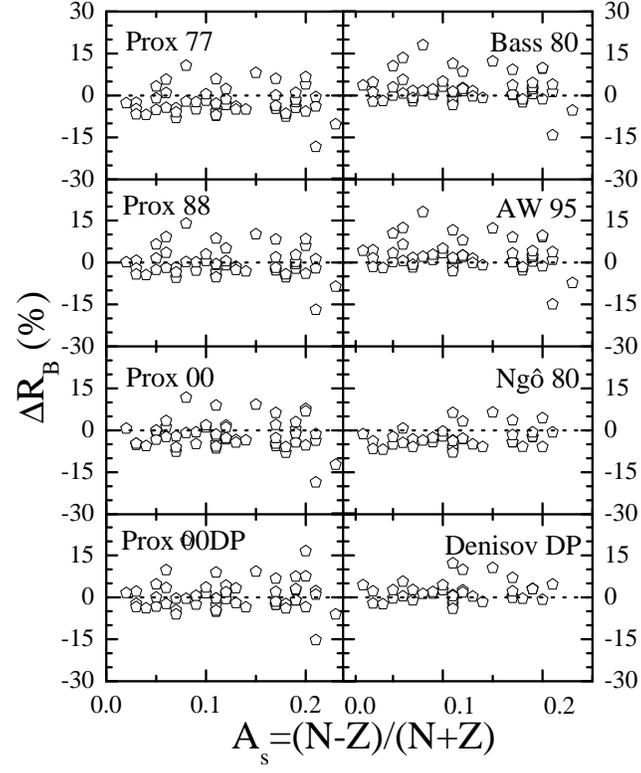}% Here is how to import EPS art
\vskip -0.50 cm \caption {Same as Fig. 5, but for percentage
difference $\Delta R_{B} (\%)$.}
\end{center}
%\vskip -4.4 cm
\end{figure}
%%%%%%%%%%%%%%%%%%%%%%%%%%%%%%%%%%%%%%%%%%%%%%%%%%%%%%%%%%%%%%
  %%%%%%%%%%%%%%%%%%%%%%%%%%%%%%%%%%Figure 5 %%%%%%%%%%%%%%%%%%%%%%%%%%
\begin{figure}[!t]
\begin{center}
\includegraphics*[scale=0.4] {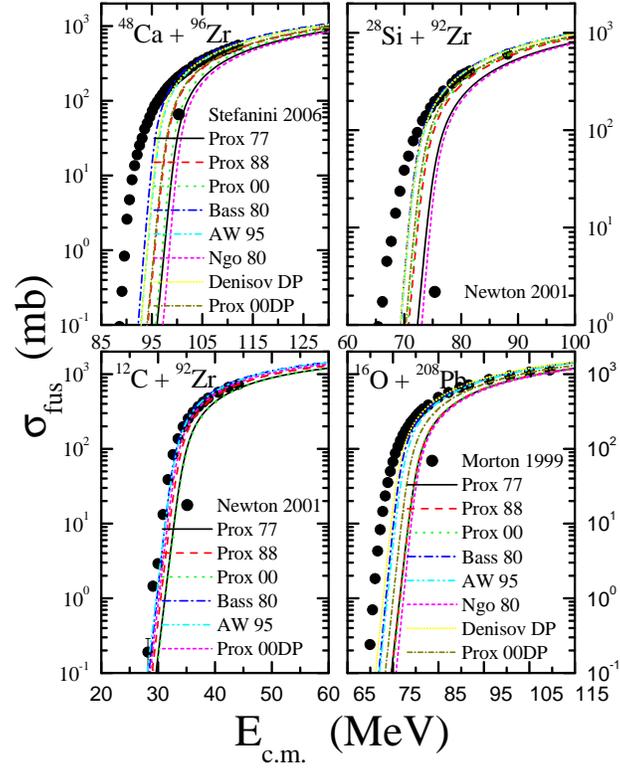}% Here is how to import EPS art
\vskip -0.50 cm \caption {(Color online) The fusion cross sections
$\sigma_{fus}$ (mb) as a function of center-of-mass energy
$E_{c.m.}~\rm (MeV)$.  For the clarity, only latest versions of
different proximity potentials are shown. The experimental data
are taken from Stefanini 2006~\cite{Stefanini06}, Newton
2001~\cite{newton01}, and Morton 1999~\cite{Morton01}.}
\end{center}
%\vskip -4.4 cm
\end{figure}
%%%%%%%%%%%%%%%%%%%%%%%%%%%%%%%%%%%%%%%%%%%%%%%%%%%%%%%%%%%%%%%%%%%%%%%%%%%%%%%%%%%%%%%%%%

%%%%%%%%%%%%%%%%%%%%%%%%%%%%%%%%%%Figure 6 %%%%%%%%%%%%%%%%%%%%%%%%%%
\begin{figure}[!t]
\begin{center}
\includegraphics*[scale=0.4] {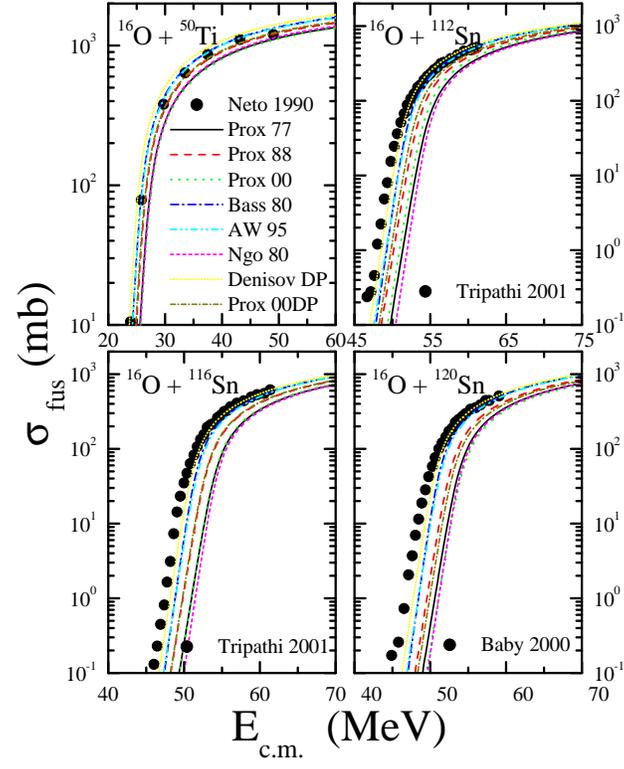}% Here is how to import EPS art
\vskip -0.50 cm \caption {(Color online) Same as Fig. 7, but for
different systems explained in the text. The experimental data are
taken from Neto 1990~\cite{Neto90}, Tripathi
2001~\cite{Vandana01}, and Baby 2000~\cite{Baby2000}. }
\end{center}
%\vskip -4.4 cm
\end{figure}
%%%%%%%%%%%%%%%%%%%%%%%%%%%%%%%%%%%%%%%%%%%%%%%%%%%%%%%%%%%%%%
\newpage
%%%%%%%%%%%%%%%%%%%%%%%%%%%%%%%%%%%%%%%%%%%%%%%%%%%%%%%%%%%%%%%%%%%%%%%%%%%%%%%%%%%%%%%%%%
\begin{table*}
\caption{\label{table}The fusion barrier heights V$_{B}$ (in MeV)
and positions R$_{B}$ (in fm) using different proximity potentials
for 60 asymmetric systems. The corresponding experimental values
are also listed. }
%\begin{ruledtabular}
% \fl
\begin{tabular}{cccccccccccc} \hline \hline
%\begin{tabular}{right-justified}
Reaction &
\multicolumn{2}{c}{Prox 77}&\multicolumn{2}{c}{Prox 88}&\multicolumn{2}{c}{Prox 00}&\multicolumn{2}{c}{Prox 00DP}&\multicolumn{2}{c}{Empirical}\\
\hline &V$_B$&R$_B$ &V$_B$&R$_B$ &V$_B$&R$_B$
&V$_B$&R$_B$&V$_B$&R$_B$ &$Ref.$
\\ \hline
%%%%%%%%%%%%%%%%%%%%%%%%%%%%%%%%%%%%%%%%%%%X  Table-1  X%%%%%%%%%%%%%%%%%%%%%%%%%%%%%%%%%%%%%%%%%%%%%%%%%%%%%%%%%%%%%%%%%%%%%%%%%%%%%%%%%
$^{7}$Li + $^{27}$Al    &6.52  &7.78   &6.34  &8.03   &6.80  &7.45   &6.34   &8.08   &7.38   &7.36&~\cite{Padron02}   \\
$^{12}$C + $^{17}$O     &8.22  &7.56   &7.98  &7.81   &8.46  &7.39   &7.93   &7.92   &8.20   &7.76 &~\cite{tighe93}  \\
$^{11}$B + $^{27}$Al    &10.68 &7.94   &10.39 &8.19   &11.09 &7.64   &10.62  &8.05   &11.20  &7.69 &~\cite{Padron02}   \\
$^{6}$Li + $^{59}$Co    &12.64 &8.41   &12.31 &8.66   &12.58 &8.49   &11.78  &9.14   &12.00  &7.60&~\cite{beck03}   \\
$^{4}$He + $^{164}$Dy   &17.71 &9.90   &17.36 &10.15  &17.36 &10.20  &16.01  &11.09  &17.14  &10.32&~\cite{vaz81}   \\
$^{4}$He + $^{209}$Bi   &21.30 &10.44  &20.89 &10.64  &20.63 &10.81  &19.20  &11.70  &20.98  &10.04 \\
& & & & & & & & & $\pm$0.05 & $\pm$0.01 &~\cite{kolata98}   \\
$^{26}$Mg + $^{30}$Si   &25.61 &8.64   &24.97 &8.89   &25.05 &8.86   &24.71  &9.01   &24.80  &9.05 &~\cite{Morsad90}   \\
$^{6}$He + $^{238}$U    &22.06 &11.22  &21.69 &11.42  &22.56 &10.97  &21.21  &11.74  &20.28  &12.50&~\cite{trotta2000}   \\
$^{6}$Li + $^{144}$Sm    &25.26 &9.80  &24.72 &10.05  &25.18 &9.85  &23.69  &10.53  &24.65  &10.20&~\cite{rath09}   \\
$^{14}$N + $^{59}$Co    &28.19 &8.83   &27.50 &9.08   &28.13 &8.87   &27.37  &9.16   &26.13  &9.60 &~\cite{gomes91}  \\
$^{7}$Li + $^{159}$Tb   &25.50 &10.20  &25.00 &10.45  &26.76 &10.15  &24.32  &10.77  &23.81  &11.03&~\cite{vaz81}   \\
$^{24}$Mg + $^{35}$Cl   &31.18 &8.60   &30.39 &8.85   &30.36 &8.90   &30.04  &8.98   &30.70  &8.84 &~\cite{Cavallera90}  \\
$^{16}$O + $^{58}$Ni   &33.32 &8.85   &32.51 &9.10   &33.52 &8.82   &32.72  &9.09   &31.67  &9.30 &~\cite{newton04}  \\
$^{18}$O + $^{64}$Ni    &32.08 &9.25   &31.35 &9.50   &32.32 &9.20   &31.58  &9.42   &32.50  &9.04 &~\cite{Silva97}   \\
$^{12}$C + $^{92}$Zr    &33.88 &9.38   &33.12 &9.63   &33.98 &9.37   &32.78  &9.79   &32.31  &9.68&~\cite{newton04} \\
$^{6}$Li + $^{208}$Pb   &31.17 &10.57  &30.59 &10.77  &31.11 &10.60  &29.49  &11.25  &30.10  &11.00&~\cite{Liu05}   \\
$^{16}$O + $^{72}$Ge    &36.79 &9.22   &35.94 &9.42   &36.80 &9.23   &35.96  &9.45   &35.40  &9.70 &~\cite{Aguilera95}   \\
$^{36}$S + $^{48}$Ca    &44.63 &9.51   &43.65 &9.76   &44.67 &9.55   &43.70  &9.78   &43.30  &&~\cite{stefanini08}   \\
$^{10}$Be + $^{209}$Bi  &40.50 &11.02  &39.78 &11.22  &40.59 &10.99  &39.11  &11.44  &37.60  &13.50 &~\cite{Kolata04}   \\
$^{19}$F + $^{93}$Nb    &50.34 &9.74   &49.24 &9.99   &49.27 &10.02  &49.27  &10.02  &46.60  &9.20   \\
& & & & & & & & &  $\pm$0.10 & $\pm$0.10 &~\cite{Prasad96}   \\
$^{12}$C + $^{152}$Sm   &48.37  &10.28  &47.41 &10.48  &48.98 &10.17  &47.60  &10.49  &46.39  &10.77&~\cite{vaz81}   \\
$^{16}$O + $^{116}$Sn   &53.56 &9.94  &52.43 &10.19  &53.48 &10.01  &52.35  &10.23  &50.94  &10.36 &~\cite{Vandana01}   \\
$^{18}$O + $^{124}$Sn   &51.99 &10.27  &50.97 &10.52  &51.89 &10.33  &50.81  &10.55  &49.30  &10.98 &~\cite{Sinha01}   \\
$^{48}$Ca + $^{48}$Ca   &53.96 &9.89   &52.84 &10.09  &53.93 &9.89   &52.86  &10.11  &51.70  &10.38 &~\cite{Trotta01}  \\
$^{27}$Al + $^{70}$Ge   &57.62 &9.59   &56.34 &9.84   &57.74 &9.58   &57.74  &9.58   &55.10  &10.20 &~\cite{Aguilera90}   \\
$^{40}$Ca + $^{48}$Ti   &61.67 &9.46   &60.27 &9.71   &60.71 &9.64   &60.71  &9.64   &58.17  &9.97 \\
& & & & & & & & & $\pm$0.62 & $\pm$0.07 &~\cite{Sonz98}  \\
$^{35}$Cl + $^{54}$Fe   &62.04 &9.46   &60.62 &9.71   &60.85 &9.66   &60.27  &9.79   &58.59  &10.14 &~\cite{Szanto90}  \\
$^{37}$Cl + $^{64}$Ni   &64.41 &9.82   &63.03 &10.07  &64.02 &9.91   &63.37  &10.05  &60.60  &10.59 &~\cite{Vega90}   \\
$^{46}$Ti + $^{46}$Ti   &67.15 &9.56   &65.64 &9.81   &66.34 &9.70   &65.38  &9.87   &63.30  &10.27 &~\cite{Stefanini02}   \\
$^{12}$C + $^{204}$Pb   &60.73 &10.84  &59.61 &11.09  &60.96 &10.85  &59.08  &11.22  &57.55  &11.34&~\cite{newton04} \\
$^{16}$O + $^{144}$Sm   &64.16 &10.31  &62.86 &10.56  &64.01 &10.38  &62.47  &10.63  &61.03  &10.85&~\cite{newton04} \\
$^{40}$Ar + $^{58}$Ni   &68.84 &9.72  &67.33 &9.97  &67.93 &9.92  &67.93  &9.92  &66.32  &10.16&~\cite{vaz81} \\
$^{37}$Cl + $^{73}$Ge   &72.43 &10.00   &70.91 &10.25  &71.88 &10.11  &70.74  &10.30  &69.20  &10.60&~\cite{quirz01}   \\
$^{28}$Si + $^{92}$Zr   &74.52 &10.00  &72.95 &10.25  &72.72 &10.30  &72.35  &10.34  &70.93  &10.19&~\cite{newton04} \\
$^{16}$O + $^{186}$W    &73.09 &10.86  &71.74 &11.06  &71.39 &11.18  &70.03  &11.40  &68.87  &11.12 &~\cite{newton04} \\
$^{48}$Ti + $^{58}$Ni   &82.70 &9.89   &80.91 &10.14  &81.34 &10.13  &81.34  &10.13  &78.80  &9.80  \\
& & & & & & & & & $\pm$0.30 & $\pm$0.30 &~\cite{Vinod96}   \\
$^{32}$S + $^{89}$Y   &82.52 &10.06  &80.78 &10.31  &81.38 &10.23  &80.62  &10.36  &77.77  &10.30&~\cite{newton04} \\
$^{36}$S + $^{90}$Zr    &82.99  &10.30 &81.30  &10.55 &82.35  &10.41  &81.10   &10.60  &79.00  &10.64 &~\cite{Stefanini2000}   \\
$^{16}$O + $^{208}$Pb   &79.38  &11.09 &77.96  &11.29 &79.30  &11.13  &77.78   &11.35  &74.90  &11.76 &~\cite{Liu05}   \\
$^{35}$Cl + $^{92}$Zr   &88.58 &10.25  &86.75 &10.50  &87.64 &10.39  &86.41  &10.56  &82.94  &10.20&~\cite{newton04} \\
$^{28}$Si + $^{120}$Sn  &89.43  &10.49 &87.65  &10.69 &88.12  &10.65  &88.12   &10.65  &85.89  &11.04 &~\cite{Baby2000}   \\
$^{19}$F + $^{197}$Au   &85.70  &11.15 &84.16  &11.35 &85.33  &11.20  &85.33   &11.20  &81.61  &11.32&~\cite{newton04} \\
$^{16}$O + $^{238}$U    &86.86  &11.39 &85.37  &11.59 &87.46  &11.30  &85.81   &11.56  &80.81  &11.45&~\cite{newton04} \\
$^{35}$Cl + $^{106}$Pd  &99.86  &10.48 &97.85  &10.68 &98.75  &10.62  &97.45   &10.74  &94.30  &11.27 &~\cite{capurro02}   \\
$^{58}$Ni + $^{60}$Ni   &102.83 &10.16 &100.67 &10.41 &102.07 &10.26  &102.07  &10.26  &96.00  &10.26 &~\cite{newton04}   \\
$^{32}$S + $^{116}$Sn   &101.78 &10.49 &99.75  &10.74 &100.65 &10.64  &99.73   &10.76  &97.36  &10.80 &~\cite{Vandana01}   \\
$^{40}$Ca + $^{90}$Zr   &103.60 &10.30 &101.46 &10.55 &102.57 &10.43  &102.10  &10.48  &96.88  &10.53&~\cite{newton04} \\
$^{48}$Ca + $^{96}$Zr   &99.33  &10.80 &97.46  &11.00 &98.73  &10.90  &97.28   &11.04  &95.90  &11.21 &~\cite{Stefanini06}   \\
$^{28}$Si + $^{144}$Sm  &108.00  &10.78 &105.90  &10.98 &105.40  &11.04  &105.03   &11.13  &103.89  &10.93 &~\cite{newton04}   \\
$^{50}$Ti + $^{93}$Nb   &112.74 &10.71 &110.54 &10.96 &111.25 &10.87  &110.38  &10.99  &106.90 &&~\cite{Stelson90}   \\
$^{40}$Ca + $^{124}$Sn   &123.11 &10.90 &120.78 &11.10 &121.55 &11.01  &121.55  &11.01  &112.93 &10.08 &~\cite{newton04}   \\
$^{28}$Si + $^{208}$Pb  &133.90 &11.56 &131.59 &11.76 &131.10
&11.79  &131.10  &11.79  &128.07 &11.45&~\cite{newton04} \\ \hline
\hline
%\begin{tabular}{right-justified}
\end{tabular}
%\end{ruledtabular}
\end{table*}
%\endgroup
%%%%%%%%%%%%%%%%%%%%%%%%%%%%%%%%%%%%%%%%%%%%%%%%%%%%%%X  Table-2  X%%%%%%%%%%%%%%%%%%%%%%%%%%%%%%%%%%%%%%%%%%%%%%%%%%%%%%%%%5

\begin{table*}
%\begin{ruledtabular}
TABLE~\ref{table}:-(continued). \\
%\emph{\label{tab:table1} continue $\dots\dots$}\\
\begin{tabular}{cccccccccccc} \hline \hline
Reaction &
\multicolumn{2}{c}{Prox 77}&\multicolumn{2}{c}{Prox 88}&\multicolumn{2}{c}{Prox 00}&\multicolumn{2}{c}{Prox 00DP}&\multicolumn{2}{c}{Empirical}\\
\hline &V$_B$&R$_B$ &V$_B$&R$_B$ &V$_B$&R$_B$
&V$_B$&R$_B$&V$_B$&R$_B$ &$Ref.$
\\ \hline
%%%%%%%%%%%%%%%%%%%%%%%%%%%%%%%%%%%%%%%%%%%%%%%%%%%%%%%%%%%%%%%%%%%%%%%%%%%%%%%%%%%%%
$^{40}$Ar + $^{165}$Ho  &141.27 &11.49 &138.78 &11.69 &138.61 &11.71  &138.61  &11.71  &141.38 &11.48 &~\cite{vaz81}   \\
$^{32}$S + $^{232}$Th  &163.08 &11.92 &160.39 &12.12 &162.32 &11.94  &160.97  &12.02  &155.73 &11.18 &~\cite{newton04}   \\
$^{40}$Ca + $^{192}$Os  &174.70 &11.71 &171.71 &11.96 &173.90 &11.74  &173.07  &11.79  &168.07 &11.05 &~\cite{newton04}   \\
$^{48}$Ti + $^{208}$Pb  &200.34 &12.18 &197.08 &12.38 &197.08 &12.34  &197.08  &12.34  &190.10 &&~\cite{Mitsuoka07}   \\
$^{56}$Fe + $^{208}$Pb  &233.61 &12.33 &229.84 &12.58 &229.74 &12.45  &229.74  &12.45  &223.00 &&~\cite{Mitsuoka07}   \\
$^{64}$Ni + $^{208}$Pb  &247.56 &12.56 &243.66 &12.76 &245.68 &12.53  &245.68  &12.53  &236.00 &&~\cite{Mitsuoka07}   \\
$^{70}$Zn + $^{208}$Pb  &262.60 &12.71 &258.53 &12.91 &259.01 &12.76  &259.01  &12.76  &250.60 &&~\cite{Mitsuoka07}   \\
$^{86}$Kr + $^{208}$Pb  &308.05 &12.99 &303.40 &13.24 &306.16 &12.92  &304.56  &12.98  &299.20 &&~\cite{Mitsuoka07}   \\
\hline \hline
\end{tabular}
\end{table*}
%%%%%%%%%%%%%%%%%%%%%%%%%%%%%%%%%%%%%%%%%%%%%%%X  Table-11  X%%%%%%%%%%%%%%%%%%%%%%%%%%%%%%%%
\begin{table*}
\caption{\label{table1}Fusion barrier heights V$_{B}$ (in MeV) and
positions R$_{B}$ (in fm) are displayed using other different
proximity potentials for 60 asymmetric systems. The limited
numbers of reactions in certain cases are due to the restriction
posed in different potentials.}
%\begin{ruledtabular}
\begin{tabular}{cccccccccccc} \hline \hline
%\begin{tabular}{right-justified}
Reaction &
\multicolumn{2}{c}{Bass 80}&\multicolumn{2}{c}{Ngo \^80}&\multicolumn{2}{c}{AW 95}&\multicolumn{2}{c}{Denisov DP}\\
\hline &V$_B$&R$_B$ &V$_B$&R$_B$ &V$_B$&R$_B$ &V$_B$&R$_B$
\\ \hline

%%%%%%%%%%%%%%%%%%%%%%%%%%%%%%%%%%%%%%%%%%%X  Table-1  X%%%%%%%%%%%%%%%%%%%%%%%%%%%%%%%%%%%%%%%%%%%%%%%%%%%%%%%%%%%%%%%%%%%%%%%%%%%%%%%%%
$^{7}$Li + $^{27}$Al    &6.20  &8.35   &-     &-     &6.31  &8.27   &-       &-  \\
$^{12}$C + $^{17}$O     &7.79  &8.13   &-     &-     &7.89  &8.10   &-       &-   \\
$^{11}$B + $^{27}$Al    &10.13 &8.50   &-     &-     &10.24 &8.49   &-       &-    \\
$^{6}$Li + $^{59}$Co    &12.00 &8.97   &-     &-     &12.14 &8.97   &-       &-     \\
$^{4}$He + $^{164}$Dy   &16.87 &10.51  &-     &-     &17.12 &10.44  &-       &-   \\
$^{4}$He + $^{209}$Bi   &20.30 &11.00  &-     &-     &20.62 &10.95  &-       &-    \\
$^{26}$Mg + $^{30}$Si   &24.33 &9.20   &25.65 &8.76  &24.42 &9.20   &23.84   &9.29     \\
$^{6}$He + $^{238}$U    &21.10 &11.83  &-     &-     &21.60 &11.59  &-       &-    \\
$^{6}$Li + $^{144}$Sm   &24.08 &10.36  &-     &-     &24.34 &10.34  &-       &-    \\
$^{14}$N + $^{59}$Co    &26.79 &9.40   &-     &-     &26.90 &9.43   &-       &-    \\
$^{7}$Li + $^{159}$Tb   &24.33 &10.76  &-     &-     &24.67 &10.72  &-       &-  \\
$^{24}$Mg + $^{35}$Cl   &29.61 &9.16   &31.19 &8.72  &29.67 &9.21   &29.21   &9.23    \\
$^{16}$O + $^{58}$Ni   &31.69 &9.41   &33.42 &8.94   &31.78 &9.44   &31.14   &9.50  \\
$^{18}$O + $^{64}$Ni    &30.53 &9.81   &32.18 &9.33  &30.70 &9.76   &29.91   &9.93    \\
$^{12}$C + $^{92}$Zr    &32.26 &9.94   &-     &-     &32.43 &9.93   &-       &-    \\
$^{6}$Li + $^{208}$Pb   &29.72 &11.14  &-     &-     &30.08 &11.11  &-       &-   \\
$^{16}$O + $^{72}$Ge    &35.02 &9.73  &36.92 &9.29   &35.14 &9.79   &34.46  &9.83  \\
$^{36}$S + $^{48}$Ca    &42.48 &10.07  &44.69 &9.59  &42.69 &10.04  &42.11   &10.09    \\
$^{10}$Be + $^{209}$Bi  &38.70 &11.59  &-     &-     &39.29 &11.48  &-       &-    \\
$^{19}$F + $^{93}$Nb    &48.01 &10.25  &50.57 &9.78  &48.24 &10.26  &47.56   &10.32     \\
$^{12}$C + $^{152}$Sm   &46.13 &10.79  &-  &-  &46.45 &10.82        &-  &-   \\
$^{16}$O + $^{116}$Sn   &51.11 &10.45  &53.85 &9.97  &51.36 &10.50  &50.61  &10.55  \\
$^{18}$O + $^{124}$Sn   &49.57 &10.83  &52.18 &10.34 &49.98 &10.80  &49.04   &10.93   \\
$^{48}$Ca + $^{48}$Ca   &51.39 &10.40  &54.06 &9.94  &51.74 &10.39  &51.13   &10.42   \\
$^{27}$Al + $^{70}$Ge   &54.97 &10.11  &57.86 &9.60  &55.13 &10.12  &54.77   &10.09     \\
$^{40}$Ca + $^{48}$Ti   &58.83 &9.97   &61.90 &9.47  &58.91 &9.99   &58.76   &9.93   \\
$^{35}$Cl + $^{54}$Fe   &59.18 &9.92   &62.28 &9.47  &59.28 &9.98   &59.11   &9.92   \\
$^{37}$Cl + $^{64}$Ni   &61.47 &10.33  &64.67 &9.87  &61.71 &10.37  &61.37   &10.33   \\
$^{46}$Ti + $^{46}$Ti   &64.10 &10.07  &67.45 &9.56  &64.21 &10.07  &64.09   &10.02   \\
$^{12}$C + $^{204}$Pb   &58.04 &11.40  &57.86 &9.60  &58.53 &11.38  &55.13   &10.12     \\
$^{16}$O + $^{144}$Sm   &61.34 &10.82  &64.59 &10.31 &61.68 &10.83  &60.85   &10.88    \\
$^{40}$Ar + $^{58}$Ni   &65.75 &10.23  &69.19 &9.71  &65.91 &10.22  &65.71  &10.20 \\
$^{37}$Cl + $^{73}$Ge   &69.19 &10.51  &72.81 &9.98  &69.48 &10.48  &69.21   &10.49   \\
$^{28}$Si + $^{92}$Zr   &71.21 &10.51  &74.96 &9.97  &71.44 &10.53  &71.32   &10.45    \\
$^{16}$O + $^{186}$W    &69.86 &11.37  &73.44 &10.85 &70.36 &11.34  &69.47   &11.45      \\
$^{48}$Ti + $^{58}$Ni   &79.08 &10.35  &83.24 &9.87  &79.28 &10.43  &79.28   &10.35    \\
$^{32}$S + $^{89}$Y     &78.91 &10.52  &83.07 &10.03  &79.15 &10.59  &79.18  &10.50 \\
$^{36}$S + $^{90}$Zr    &79.40 &10.76  &83.56 &10.26 &79.82 &10.76  &79.54   &10.73   \\
$^{16}$O + $^{208}$Pb   &75.92 &11.60  &79.76 &11.07 &76.52 &11.60  &75.55   &11.66    \\
$^{35}$Cl + $^{92}$Zr   &84.76 &10.71  &89.22 &10.16  &85.09 &10.71  &85.11  &10.65 \\
$^{28}$Si + $^{120}$Sn  &85.56 &10.95  &90.04 &10.39 &85.94 &10.93  &85.98   &10.86    \\
$^{19}$F + $^{197}$Au   &82.04 &11.66  &86.19 &11.07 &82.76 &11.60  &81.92   &11.67   \\
$^{16}$O + $^{238}$U    &83.12  &11.90  &87.20  &11.37  &83.85  &11.88  &82.77   &11.98    \\
$^{35}$Cl + $^{106}$Pd  &95.67  &10.89  &100.71 &10.38  &96.09  &10.92  &96.24   &10.81    \\
$^{58}$Ni + $^{60}$Ni   &98.53  &10.56  &103.77 &10.02  &98.80  &10.61  &99.09   &10.53    \\
$^{32}$S + $^{116}$Sn   &97.53  &10.95  &102.68 &10.39  &97.93  &10.98  &98.18   &10.85   \\
$^{40}$Ca + $^{90}$Zr   &99.28  &10.71  &104.55 &10.15  &99.58  &10.75  &99.93   &10.66   \\
$^{48}$Ca + $^{96}$Zr   &95.05  &11.26  &100.03 &10.74  &95.87  &11.23  &95.61   &11.19    \\
$^{28}$Si + $^{144}$Sm  &103.60 &11.19 &109.04  &10.61 &104.12  &11.20  &104.31   &11.12  \\
$^{50}$Ti + $^{93}$Nb   &108.10 &11.17  &113.83 &10.59  &108.77 &11.13  &108.87  &11.06   \\
$^{40}$Ca + $^{124}$Sn  &118.07 &11.31  &124.29 &10.73  &118.66 &11.31  &119.36  &11.14   \\
$^{28}$Si + $^{208}$Pb  &128.56 &11.97  &135.03 &11.37  &129.53 &11.92  &130.06  &11.79   \\
$^{40}$Ar + $^{165}$Ho  &135.70 &11.90  &142.75 &11.29  &136.97 &11.86  &137.38  &11.73   \\
$^{32}$S + $^{232}$Th   &156.86 &12.28  &164.65 &11.67  &158.17 &12.25  &-       &-    \\
$^{40}$Ca + $^{192}$Os  &168.22 &12.07  &176.93 &11.45  &169.44 &12.05  &171.15  &11.82    \\
$^{48}$Ti + $^{208}$Pb  &193.15 &12.49  &203.09 &11.86  &195.26 &12.44  &196.99  &12.15    \\
$^{56}$Fe + $^{208}$Pb  &225.72 &12.59  &237.53 &11.89  &228.16 &12.52  &230.95  &12.18    \\
$^{64}$Ni + $^{208}$Pb  &239.28 &12.81  &251.83 &12.10  &242.40 &12.68  &245.24  &12.31   \\
$^{70}$Zn + $^{208}$Pb  &253.99 &12.92  &267.37 &12.20  &257.54 &12.78  &260.75  &12.37   \\
$^{86}$Kr + $^{208}$Pb  &298.65 &13.15  &-      &-      &303.25 &12.98  &308.13  &12.32   \\
\hline \hline

%\begin{tabular}{right-justified}
\end{tabular}
%\end{ruledtabular}
\end{table*}
%%%%%%%%%%%%%%%%%%%%%%%%%%%%%%%%%%%%%%%%%%%%%%%%%%%%%%%%%%%%%%%%%%%%%%%%%%%%%%%%%%%%%%%

\end{document}